\documentclass[prl,aps,showpacs,twocolumn,superscriptaddress]{revtex4}
\usepackage{latexsym,graphicx,amsmath,amsfonts,subfigure}
\def\vec#1{{\bf #1}}
\def\ket#1{|#1\rangle}
\def\bra#1{\langle#1|}

\begin{document}
\title{Efficient and robust adiabatic universal quantum computation using STIRAP on a qubit chain}

\author{Toshio Ohshima}
\address{Institute of Industrial Science,
University of Tokyo,
4-6-1 Komaba, Meguro-ku, Tokyo 153-8505 Japan.}
\date{Received: \today}

\begin{abstract}
  It is shown that efficient and robust universal quantum computation is possible using the stimulated Raman adiabatic passage with a qubit chain as a pointer register.
\end{abstract}


\maketitle

A quantum computer (QC) can solve some class of problems more efficiently than any classical computers can, if we can implement reliable quantum gates and measurements physically. In general, quantum gates rotate state vectors through some axis by some angles. The direction of axis and the amount of angle is determined by external control signals. The requirements for the accuracy of control signals is very hard in order to fulfill the threshold criterion for fault-tolerant quantum computation.

In ref ~\cite{Ohshima2007}, we proposed a robust method of state transfer and single qubit gate using adiabatic passage along spin chains~\cite{Eckert2007, Bose2008}. In this proposal, such quantum operations are done using counter-intuitive pulse sequence which couples neighboring spins. This technique is derived from a traditional method of optical pumping called the stimulated Raman adiabatic passage (STIRAP). The fidelity of operation can be made arbitrarily good by the proper preliminary calibration of coupling Hamiltonian. However, for universal quantum computation, two qubit gates must also be made robust. In this paper, we propose a new method for doing arbitrary sequence of (single or two qubit) quantum gates with robustness. The key ingredient is again the adiabatic passage. This method is basically a variation of the adiabatic quantum computation~\cite{Farhi2000}. However, we must point out that it uses the zero energy dark state rather than the ground state. Of course in both cases adiabaticity requires that change of external parameters must be sufficiently slow to minimize nonadiabatic transition from the eigenstate to nearest eigenstates~\cite{Kato1951}. However, in our scheme the energy gap between the dark state and neighboring states below and above scales ~1/n, where n is the number of gates for computation. This means the computation time increases only linearly with the number of gates. Of course, the computation is deterministic. The concept of our method is sufficiently general and can be implemented in various physical systems, e.g., optical lattices, superconductors, and quantum dots. However, as an example, we describe an implementation using a system of spin 1/2.

We define following Hamiltonian system which acts as QC.
\begin{eqnarray}
H(s)&=&(1-s)H_{init}+sH_{final}\\
H_{init}&=&J(\ket{n+2}_c\bra{n+1}
+\ket{n+1}_c\bra{n+2})\notag\\
&+&M(\sum_{i=1}^{n}U_i\ket{i+1}_c\bra{i}+U_i^+\ket{i}_c\bra{i+1})\\
H_{final}&=&J(\ket{1}_c\bra{0}
+\ket{0}_c\bra{1})\notag\\
&+&M(\sum_{i=1}^{n}U_i\ket{i+1}_c\bra{i}+U_i^+\ket{i}_c\bra{i+1}),
\end{eqnarray}
where subscript c stands for the state of the counter (qubit chain of length n+2) and $U_i$ is i-th unitary gate of quantum computation on the N-qubit computation register.
We make M much larger than J.
Initial state (at s=0) of the QC is
\begin{eqnarray}
\ket{s=0}_{QC}=\ket{0}_c\ket{\phi}_r,
\end{eqnarray}
where $\ket{\phi}_r$ is the state encoding the arbitrary input data of computation.
We increases s from 0 to 1 slowly. The final state (at s=1) is
\begin{eqnarray}
\ket{s=1}_{QC}=\ket{n+2}_c\prod_{i=1}^nU_i\ket{\phi}_r
\end{eqnarray}

At the general value of s $(0<s<1)$, the state of QC is
\begin{eqnarray}
\ket{s}_{QC}&=&(1-s)J\ket{0}_c\ket{\phi}_r\notag\\
&+&\frac{J^2}{M}s(1-s)\sum_{i=1}^{n/2}(-1)^i\ket{2i}_c\prod_{k=1}^{2i}U_k\ket{\phi}_r\notag\\
&+&sJ\ket{n+2}_c\prod_{k=1}^{n}U_k\ket{\phi}_r.
\end{eqnarray}

Note that we need to modulate the strength of couplings only for first and last bonds of the counter chain. Compare above method with other schemes of QC with a counter~\cite{Feynman1986, Kitaev2002, Aharonov2004, Lloyd2008}.

All above results come from the adiabatic approximation. The validity of the approximation requires that change of external parameter s must be slow and proportional to the energy gap. Contrary to the conventional adiabatic quantum computations, relevant gaps are between zero energy dark state and upper and lower nearest eigenstates. It is straightforward to obtain that the gap in our system to be inverse proportional to the number of qubits in the counter chain and thus inverse proportional to the number of gates. The linear dependence of the gap on the inverse number of gates comes from the sine (or linear) dependence of eigenvalue of the Hamiltonian around zero energy. Of course there are $2^N$ degeneracy in which all possible register states belong to a same level of the counter among $2^n$ levels in total. However, this degeneracy does not contribute to the reduction of adiabaticity.
The advantage of this scheme is evident; (1) All unitary operations are embedded preliminarily in the designed Hamiltonian and do not suffer temporal fluctuation which comes from noises in controller system. We need only to prepare accurate Hamiltonian and an initial state. We can even calibrate Hamiltonian until it reaches a satisfactory level of accuracy. (2) Operation time is proportional to the number of gates, similar to standard quantum computation using circuit model. (3) After s has reached 1, answer can be obtained deterministically by measuring the register at any time, since the amplitude of the state encoding correct answer is unity. There is no need of precise timing of measurement and the repetition of computation to get a correct answer.

Next we describe the implementation of our scheme using spin degree of freedom. The QC is described by
\begin{eqnarray}
H(s)&=&(1-s)H_{init}+sH_{final}\\
H_{init}&=&\frac{J}{2}(X_{n+2}X_{n+1}+Y_{n+2}Y_{n+1})_c\notag\\
&+&\frac{M}{2}\sum_{i=1}^{n}[H_i^{r,s}(X_{i}X_{i+1}+Y_{i}Y_{i+1})_c\notag\\
&+&H_i^{r,a}(X_{i}Y_{i+1}-Y_{i}X_{i+1})_c]\\
H_{final}&=&\frac{J}{2}(X_{0}X_{1}+Y_{0}Y_{1})_c\notag\\
&+&\frac{M}{2}\sum_{i=1}^{n}[H_i^{r,s}(X_{i}X_{i+1}+Y_{i}Y_{i+1})_c\notag\\
&+&H_i^{r,a}(X_{i}Y_{i+1}-Y_{i}X_{i+1})_c],
\end{eqnarray}
where $X_i, Y_i, Z_i$ are Pauli spin matrices for i-th spin of the counter.

The symmetric and antisymmetric part of Hamiltonian for the register are
\begin{eqnarray}
H_i^{r,s}&=&\frac{1}{2}(U_i+U_i^+)\\
H_i^{r,a}&=&\frac{i}{2}(U_i-U_i^+).
\end{eqnarray}
For example, Hamiltonian for typical and useful quantum gates are
\begin{eqnarray}
&&H^s(Hadamard)=\frac{1}{\sqrt{2}}(X+Z), \\
&&H^a(Hadamard)=0,\\
\notag\\
&&H^s(\frac{\pi}{8})=\frac{1+\sqrt{2}}{2}I+\frac{1-\sqrt{2}}{2}Z, \\
&&H^a(\frac{\pi}{8})=\frac{1}{\sqrt{2}}(Z-I),\\
\notag\\
&&H^s(R_\vec{n}(\theta))=cos{\frac{\theta}{2}}I, \\
&&H^s(R_\vec{n}(\theta))=sin{\frac{\theta}{2}}(n_xX+n_yY+n_zZ),\\
\notag\\
&&H^s(CNOT)=I_1I_2+Z_1I_2+I_1X_2-Z_1X_2, \\
&&\notag\\
&&and\notag\\
&&\notag\\
&&H^a(CNOT)=0.
\end{eqnarray}
\\

Thus, we need four-spin interactions at most. A limiting case of interest is when M is larger than J in many orders of magnitude. In this case, there created almost no amplitude on the intermediate qubits in the counter, and so in the intermediate circuit nodes of quantum computation. This situation looks as if quantum computation is performed without experiencing any intermediate states.

We thank Akihito Soeda and Mio Murao for stimulating discussion on this topic.

\end{document}